\documentstyle[11pt]{article}

\newcommand{\sect}[1]{\setcounter{equation}{0}\section{#1}}
\newcommand{\subsect}[1]{\subsection{#1}}
\newcommand{\subsubsect}[1]{\subsubsection{#1}}

\def\be{\begin{equation}}
\def\ee{\end{equation}}
\def\bea{\begin{eqnarray}}
\def\eea{\end{eqnarray}}

\def\co{\Delta}  
\def\h{h}  
\def\Gco{\bar{\Delta}}

\def\b{b}

\def\J{{\tilde J}} 
\def\JJ{{\bar J}} 

\def\>#1{{\bf #1}}                 
\def\Iz#1#2{I_{\{\frac12\}}[#2\,#1]}                 
\def\Sz#1{\frac{\sinh(z\,#1)}{z}}                 
\def\1{\'{\i}}                           
\def\R{\rm I\kern-.2em R} 
\def\C{\rm I\kern-.5em C} 
\def\back{\!\!\!\!\!\!}

\def\tfrac#1#2{ {\scriptstyle { \frac {#1}{#2}}}}         
\def\pois#1#2{\left\{ {#1},{#2} \right\}}

\begin{document}
\thispagestyle{empty}
\hfill\today 
\vspace{1.5cm}

\begin{center}

{\LARGE{\bf{$N=2$ Hamiltonians}}} 
\smallskip
{\LARGE{\bf{with $sl(2)$ coalgebra symmetry}}} 
\smallskip
{\LARGE{\bf{and their integrable deformations}}} 

\vspace{1.4cm} 

ANGEL BALLESTEROS\vspace{.2cm}\\
{\it Departamento de F\1sica, Universidad de Burgos\\
Pza. Misael Ba\~nuelos s.n., 09001-Burgos, Spain} 
\vspace{.3cm}

ORLANDO RAGNISCO\vspace{.2cm}\\
{\it Dipartimento di Fisica,  Universit\'a di Roma TRE\\
Via della Vasca Navale 84, 00146-Roma, Italy} 
\vspace{.3cm}

\end{center} 
  
\bigskip

\begin{abstract} 

Two dimensional classical integrable systems and different
integrable deformations for them are derived from phase space
realizations of classical $sl(2)$ Poisson coalgebras and their $q-$deformed 
analogues. Generalizations of Morse,
oscillator and centrifugal potentials are obtained. The $N=2$ Calogero
system is shown to be $sl(2)$ coalgebra invariant and the well-known
Jordan-Schwinger realization can be also derived from a
(non-coassociative) coproduct on $sl(2)$. The Gaudin Hamiltonian
associated to such Jordan-Schwinger construction is presented. Through
these examples, it can be clearly appreciated how the coalgebra
symmetry of a hamiltonian system allows a straightforward construction
of different integrable deformations for it. 

\end{abstract} 

\bigskip

%%%%%%%%%%%%%%%%% INTRODUCTION %%%%%%%%%%%

\sect{Introduction}

A Poisson coalgebra $(A,\co)$ is a Poisson algebra $A$ endowed with a
Poisson map $\co$ between $A$ and $A\otimes A$. In other words, if we
assume that $A$ is the dynamical algebra for a one-particle problem,
the coproduct $\co$ provides a two-particle realization of the same
dynamical symmetry. In fact, under certain conditions the coproduct
$\co$ can be uniquely generalized to a Poisson map between $A$ and
$A\otimes A\otimes \dots ^{N)}\otimes A$, and the $N$-particle
realization of the symmetry arises. By following this approach, a
systematic construction of $N$-dimensional completely integrable
 Hamiltonians from any coalgebra with Casimir
element $C$ has been introduced in \cite{BR}. Moreover, since $q-$ 
deformations (see, for instance, \cite{KR}-\cite{GRAS}) can be
translated in a classical mechanical context as deformations of Poisson
algebras preserving a coalgebra structure, such construction can be
applied for them and leads to a systematic derivation  of integrable
deformations of Hamiltonian systems.

The aim of the present paper is two-fold. On one hand, we use the
$sl(2)$ Poisson coalgebra and its deformations to provide new classical
mechanical examples of this general construction. In Section 2 we
recall such  Poisson
$sl(2)$ coalgebras; in Section 3 the general construction is reviewed
through an example based on the Gelfan'd-Dyson realization \cite{GD}
of $sl(2)$ and in Section 4 new $N=2$ integrable systems related to
Morse, oscillator and centrifugal problems are given. Another
interesting example of two-dimensional coalgebra symmetry is provided
by the
$N=2$ Calogero system \cite{Cal74}, that is shown to be canonically
equivalent to one of the systems that have been previously derived.
Therefore, the results here presented can be used to construct two new
different integrable deformations of this remarkable Hamiltonian.

As a
general fact, $q-$deformations introduce hyperbolic functions of
the canonical variables, and the associated integrals of the motion
have also such kind of hyperbolic terms depending on positions and/or
momenta. Note also that the existence of the
$N$-dimensional generalization of all these systems is ensured
(although we will not describe their explicit form here) and that phase
space realizations of coalgebras play an essential role in the
formalism. 

Secondly, in Section 5 we present the generalization of this
construction to systems in which the dynamical symmetry algebra $A$ is
given in terms of an elementary canonical realization that already
contains two pairs of  conjugated variables. By using the classical
Jordan-Schwinger (JS) realization
\cite{jord,Sch} as an outstanding axample of this kind of situation, we
show how complete integrability is also preserved for the corresponding
system, now with
$2\,N$ degrees of freedom. Consequently, the deformation of such systems
poses the interesting problem of deforming the JS map. It turns out that
coalgebra symmetry is also essential at this point, since we show that
the Jordan-Schwinger (JS) realization of
$sl(2)$ is canonically equivalent to a
reducible representation of $sl(2)$ given by a non-coassociative
coproduct. From it, a standard deformation of the JS realization is
given and related
$2\,N$ dimensional integrable systems can be defined.

\sect{$sl(2)$ Poisson coalgebras}

Let us consider classical angular momentum variables $J_3,J_\pm$
and the associated $sl(2)$ Poisson-Lie algebra 
\be
\pois{J_3}{J_\pm}=\pm\,2\,J_\pm, \qquad 
\pois{J_+}{J_-}=J_3. \qquad 
\label{hg}
\ee
with Casimir function
\be
C(J_3,J_+,J_-)=\tfrac{1}{4} J_3^2 +J_+J_-.
\label{hgb}  
\ee
The Poisson algebra (\ref{hg}) is endowed with a coalgebra structure by
the usual ``primitive" coproduct defined between $sl(2)$ and
$sl(2)\otimes sl(2)$
\be
\co(J_3)=1\otimes J_3 + J_3\otimes 1,\qquad
\co(J_\pm)=1\otimes J_\pm + J_\pm\otimes 1.
\label{prcop}
\ee
Compatibility between (\ref{prcop}) and (\ref{hg}) means that $\co$ is
a Poisson map: the three functions defined through (\ref{prcop}) close
also a Poisson $sl(2)$ algebra.

There are few deformations of the $sl(2)$ algebra for which there exist
compatible deformations of the coproduct (\ref{prcop}). In fact, two
relevant and distinct structures appeared in quantum group literature
during last years, and they can be realized as Poisson algebras as
follows:

\noindent $\bullet$ The {\it ``standard"} deformation \cite{KR,Dr,Ji},
$sl_q(2)$
$(q=e^z)$ given by the following deformed Poisson brackets
\be
\pois{\J_3}{\J_+}=2\,\J_+,\quad \pois{\J_3}{\J_-}=-2\,\J_-,\quad 
\pois{\J_+}{\J_-}=\frac{\sinh (z\J_3)}{z},
\label{lc} 
\ee
which are compatible with the deformed coproduct
\bea
&& \co_z(\J_3) =1 \otimes \J_3 + \J_3\otimes 1,\cr  
&&  \co_z(\J_+) =e^{-\tfrac{z}{2}\J_3} \otimes \J_+ + \J_+\otimes
e^{\tfrac{z}{2}\J_3}; \label{lb} \\
&&  \co_z(\J_-) =e^{-\tfrac{z}{2}\J_3} \otimes \J_- + \J_-\otimes
e^{\tfrac{z}{2}\J_3};\nonumber
\eea
in the sense that that (\ref{lb}) is a Poisson algebra
homomorphism with respect to (\ref{lc}). The function
\be
C_z(\J_3,\J_\pm)=\left(\frac{\sinh (\tfrac{z}{2} \J_3)}{z}\right)^2 +
\J_+\,\J_-, \label{le}
\ee 
is the deformed Casimir for this Poisson coalgebra.

\noindent $\bullet$ The {\it ``non-standard"} deformation
\cite{Demidov,Zakr,Ohn},
$sl_\h(2)$ whose defining relations are
\be
\pois{\J_3}{\J_+}=2\,\frac{\sinh (\h\J_-)}{\h},\quad
\pois{\J_3}{\J_-}=-2\,\J_-\cosh (\h\J_+),\quad 
\pois{\J_+}{\J_-}=\J_3.
\label{ls} 
\ee
\bea
&& \co_\h(\J_+) =1 \otimes \J_+ + \J_+\otimes 1,\cr  
&&  \co_\h(\J_-) =e^{- \h\J_+} \otimes \J_- + \J_-\otimes
e^{ \h\J_+}; 
\label{lr}\cr
&&  \co_\h(\J_3) =e^{- \h\J_+} \otimes \J_3 + \J_3\otimes
e^{ \h\J_+}; 
\eea
The non--standard deformed Casimir is now
\be
C_h(\J_3,\J_+,\J_-)=\tfrac{1}{4} \J_3^2 +\frac{\sinh (\h\J_+)}{\h}\,\J_-
\label{lu}  
\ee
Equivalently, an isomorphic non-standard Poisson deformation is obtained
by choosing $\J_-$ as the primitive generator. We then find
that
\be
\pois{\J_3}{\J_+}=2\,\J_+\cosh (\h\J_-),\quad
\pois{\J_3}{\J_-}=-2\frac{\sinh (\h\J_-)}{\h},\quad 
\pois{\J_+}{\J_-}=\J_3.
\label{lsm} 
\ee
\bea
&& \co_\h(\J_-) =1 \otimes \J_- + \J_-\otimes 1,\cr  
&&  \co_\h(\J_+) =e^{- \h\J_-} \otimes \J_+ + \J_+\otimes
e^{ \h\J_-}; 
\label{lrr}\cr
&&  \co_\h(\J_3) =e^{- \h\J_-} \otimes \J_3 + \J_3\otimes
e^{ \h\J_-}; 
\eea
The non--standard deformed Casimir is now
\be
C_h(\J_3,\J_+,\J_-)=\tfrac{1}{4} \J_3^2 +\J_+\,\frac{\sinh
(\h\J_-)}{\h}
\label{lux}  
\ee

In what follows we shall make use of these coalgebra
symmetries in order to construct new integrable systems. 
Note that the undeformed $sl(2)$ structure is smoothly recovered when
deformation parameters vanish. Jacobi identity can be also easily
checked for (\ref{lc}),(\ref{ls}) and (\ref{lsm}).

\sect{Gelfan'd-Dyson map, Gaudin magnet and deformations}

Let us now recall the general construction \cite{BR} through an example
provided by the classical phase space analogue of the one-boson
(polynomial) Gelfan'd-Dyson (GD) realization of $sl(2)$:
\be
J_3=2\,p\,q - \b,\qquad
J_+=p,\qquad
J_-=-p\,q^2 + \b\,q,
\label{hh}
\ee
where $\b$ is a real constant that labels the representation through
the Casimir (\ref{hgb}), which turns out to be $b^2/4$ under (\ref{hh}).

Let us now consider an {\it arbitrary} function ${\cal H}(J_3,J_\pm)$.
Since the coproduct (\ref{prcop}) is an algebra homomorphism, it is
immediate to prove that the two-particle Hamiltonian that can be defind
through the coproduct of $H$ in the form
\be
H^{(2)}=\co({\cal H}(J_3,J_\pm))={\cal H}(\co(J_3),\co(J_\pm)),
\label{ff}
\ee
will commute with the coproduct $C^{(2)}$ of the Casimir element:
\be
\pois{H^{(2)}}{C^{(2)}}=\pois{\co({\cal
H})}{\co(C)}=\co\left(\pois{{\cal H}}{C}
\right)=0.
\label{prove}
\ee

Therefore, {\it any} function of the generators defines a two-site
integrable Hamiltonian. In particular, the Casimir itself can be taken
as the function ${\cal H}$. In that case, the other integral of motion
will be given by the coproduct of any of the generators of the algebra.
Explicitly, in the $sl(2)$ case we have
\bea
&& H_C^{(2)}:=C(\co(J),\co(J_+),\co(J_-))\cr
&&\qquad =\tfrac{1}{4}(1\otimes J_3 + J_3\otimes 1)^2 + (1\otimes J_+ +
J_+\otimes 1)(1\otimes J_- + J_-\otimes 1)\cr
&&\qquad =1\otimes C + C\otimes 1 + \tfrac12\,J_3\otimes J_3
+J_-\otimes J_+ +J_+\otimes J_-, 
\label{hi}
\eea
which is just the two-site Gaudin magnet \cite{Gau,EEKT,Cal}.
Once (\ref{hi}) is realized in terms of two copies of (\ref{hh}) we
shall obtain the integrable two-particle Hamiltonian
\bea
&& H_C^{(2)}(q_1,q_2,p_1,p_2):= \tfrac{b^2}{4} + \tfrac{b^2}{4} +
\tfrac{1}{2}
\,(2\,p_1\,q_1 - b)(2\,p_2\,q_2
- b) \cr
&&\qquad\qquad\quad + (-p_1\,q_1^2 + b\,q_1)\,p_2 + p_1\,(-p_2\,q_2^2 +
b\,q_2)
\cr
&&\qquad\qquad\quad = -\,p_1\,p_2\,(q_1-q_2)^2 -
b\,(p_1-p_2)\,(q_1-q_2) + b^2. 
\label{hj}
\eea
Since $H_C^{(2)}$ is the coproduct of $C$, it will commute with, for
instance, with
$\co(J_+)$, which is just the total momentum $p_1+p_2$ and gives us the 
additional
integral of the motion. The generalization of this result to an
$N$-site Gaudin magnet is straightforward by taking into account
the appropriate $N$-th generalization of the coproduct. By following
\cite{BR}, we obtain that
\bea
&&H^{(N)}=\sum_{i=1}^{N}{C_i} + \sum_{i<j}^{N}{(\tfrac12\,J_3^i\, J_3^j
+J_-^i\, J_+^j +J_+^i\, J_-^j)}\label{gmg}\\
&&\qquad\qquad=\sum_{i<j}^{N} \left\{ -\,p_i\,p_j\,(q_i-q_j)^2 -
b\,(p_i-p_j)\,(q_i-q_j)\right\} + \tfrac{b^2}{4}\,N^2.
\label{hk}
\eea
The $m=2,\dots,N$
hamiltonians $H^{(m)}$, together with the total momentum
$\co^{(N)}(J_+)=p_1+p_2+\dots+p_N$ are again $N$ functionally
independent
integrals of  motion in involution. Note also that we could have
taken a different realization on each lattice site through different
$b_i$ constants in (\ref{hh}), and the formalism will guarantee the
integrability in the same manner (this kind of realizations will be
relevant when the coalgebra symmetry of the Calogero system is analysed
in Section 4).

Now, the same construction can be applied to deformed $sl(2)$
coalgebras. We shall consider the non-standard one (\ref{ls}-\ref{lu}).
The following deformed phase-space realization of (\ref{ls}) can be
found
\bea
&& \J_+=p,\qquad \J=2\,\frac{\sinh (\h\,p)}{\h}\,q - b\,\cosh (\h\,p),
\cr
&& \J_-=-\frac{\sinh (\h\,p)}{\h}\,q^2 + b\,\cosh (\h\,p)\,q - b^2\,
\tfrac{\h}{4}\,\sinh (\h\,p),
\label{lt}
\eea
which leads again the same $b^2/4$ constant when substituted in the
deformed Casimir function (\ref{lu}). Therefore, an integrable
deformation of the system (\ref{hj}) will be given by the deformed
coproduct of the (also deformed) Casimir (\ref{lu}). In terms of two
phase space realizations of the type (\ref{lt}), the coproduct
(\ref{lr}) defines the two-particle functions
\bea
&& \tilde{f}_+=\co_\h(\J_+) =p_1 + p_2,\cr  
&& \tilde{f}_-= \co_\h(\J_-) =e^{- \h\,p_1} \left\{-\frac{\sinh
(\h\,p_2)}{\h}\,q_2^2 + b\,\cosh (\h\,p_2)\,q_2 - b^2\,
\tfrac{\h}{4}\,\sinh (\h\,p_2) \right\}\cr
&& \qquad\qquad\qquad\qquad\qquad +
\left\{-\frac{\sinh
(\h\,p_1)}{\h}\,q_1^2 + b\,\cosh (\h\,p_1)\,q_1 - b^2\,
\tfrac{\h}{4}\,\sinh (\h\,p_1) \right\}\, e^{ \h\,p_2}; 
\label{lrpq}\cr
&& \tilde{f}_3=  \co_\h(\J_3) =
e^{- \h\,p_1} \left\{ 2\,\frac{\sinh (\h\,p_2)}{\h}\,q_2 - b\,\cosh
(\h\,p_2)\right\}\cr
&& \qquad\qquad\qquad\qquad\qquad +
 \left\{ 2\,\frac{\sinh (\h\,p_1)}{\h}\,q_1 - b\,\cosh
(\h\,p_1)\right\}
\,e^{ \h\,p_2}; 
\eea

The corresponding Hamiltonian is
obtained as
\bea
&&\!\!\!\!\!\!\!\!\!\!\!\!
{H_\h}^{(2)}(q_1,q_2,p_1,p_2)=\co_\h\left( \tfrac{1}{4} \J_3^2 +
\frac{\sinh (\h\J_+)}{\h}\,\J_-\right)=\tfrac{1}{4} (\tilde{f}_3)^2
+\frac{\sinh (\h\tilde{f}_+)}{\h}\,\tilde{f}_-\cr
&&
\!\!\!\!\!\!\!\!\!\!\!\!
=-\frac{\sinh \h p_1}{\h}e^{-\h (p_1-p_2)}
\,\frac{\sinh \h p_2}{\h}\,(q_1-q_2)^2
\cr &&
\quad -
b\,\frac{1-e^{-2\h(p_1-p_2)}}{2\h}\,(q_1-q_2) 
+ \tfrac{b^2}{4}\left(1+ e^{-2\h\,p_1} + e^{2\h\,p_2} + 
e^{-2\h(p_1-p_2)}
\right).
\label{lw}
\eea
This Hamiltonian will commute, by construction,  with the coproduct  of
$\J_+$ (i.e., the total momentum $p_1+p_2$ again) and the limit
$\h\to 0$ of this expression leads to (\ref{hj}). The $N$-th dimensional
integrable generalization of this system is given by the $N$-th
coproduct of the deformed Casimir, and the integrals of motion will be
the lower degree coproducts of such Casimir and the $N$-th
total momenta. Apart from the GD realization (\ref{hh}), the essential
ingredients for the explicit formulation of such system can be
extracted from \cite{BR}. In what follows, we shall restrict our study
to the construction of two-body problems. However, we stress that, due
to the underlying coalgebra symmetry, all the systems constructed in
this way will have
$N$-dimensional integrable generalizations.

\sect{$N=2$ systems and their integrable deformations}

Let us now use realizations of $sl(2)$ linked to well-known dynamical
symmetries of physically relevant potentials like the Morse and the
oscillator with centrifugal term in order to obtain new systems with
$sl(2)$ coalgebra symmetry.

\subsect{Morse potential realization}

If we consider the following $sl(2)$ phase space realization 
\bea
&& J_3= 2\, p_1  ,\cr
&& J_+= \tfrac12  e^{- q_1}  ,\cr
&& J_-= - 2\, p_1^2\, e^{q_1} - a_1\, e^{q_1}   ,
\label{morser}
\eea
The dynamical Hamiltonian
\be
H_m=\tfrac18 J_3^2 - 4\,s\,J_+ + 4\,s\,J_+^2
\label{morseh}
\ee
leads to the Morse one:
\be
H_m=\tfrac12 p_1^2 + s\,(e^{-2\,q_1} -2\,e^{-q_1}).
\ee
The corresponding Casimir is
\be
C_m=\tfrac12 J_3^2 + J_+\,J_-=-a/2.
\ee

The two-body system is obtained by applying the method described above.
The coproduct of the Hamiltonian (\ref{morseh}) in the realization 
(\ref{morser}) is
\be
H_m^{(2)}=\tfrac12 (p_1+p_2)^2 + s\,(e^{-2\,q_1} -2\,e^{-q_1}) + 
s\,(e^{-2\,q_2} -2\,e^{-q_2}) + 2\,s\,e^{-(\,q_1+q_2)}.
\label{morseh2}
\ee
Note that we can take different realizations on each copy of the
algebra ($a_1$ and $a_2$ have not to be the same). Therefore, the
coproduct of the Casimir gives a two-parametric family of constants of
the motion in the form:
\be
C_m^{(2)}=-\tfrac12 (a_1+a_2) -(\tfrac12 a_1 + p_1^2)\,e^{\,q_1 - q_2}
-(\tfrac12 a_2 + p_2^2) e^{-(q_1 - q_2)} + 2\,p_1\,p_2. 
\label{morsessd}
\ee

%%%%%%%%%%%%%%%%%%%%%%%%%%%%%%%%%%%%%%%%%%%%%%%%%%%%%%%%%%%%%%%%

The Hamiltonian (\ref{morseh2}) can be diagonalized if we consider the
canonical transformation
\be
P_1=p_1+p_2,\qquad P_2=p_2,\qquad Q_1=q_1 \qquad Q_2=q_2 - q_1.
\label{cantras}
\ee
This
leads to a Hamiltonian in which
$P_2$ does not appear, and consequently the relative
position between both particles $Q_2=q_2-q_1$ is a constant of
the motion. Namely
\be
H_{Q_2}^{(2)}=\tfrac12 P_1^2 +
s(e^{-2\,Q_1}\,(1+e^{-Q_2})^2 - 2\,e^{-Q_1}\,(1+e^{-Q_2})).
\label{morsered}
\ee
 Note that
(\ref{morsered}) is a Morse-type problem depending on the constant
parameter
$Q_2$. Actually, in the limit $Q_2\to\infty$ we recover the original
Morse potential. It is worth recalling that this kind of
parameter-dependent dynamics was already observed in \cite{LMSV}, where
the classical motion on the Poisson-Lie $sl(2)$ group was considered.
On the other hand, it is immediate to check that the $N$-dimensional
generalization of the system (\ref{morseh2}) can be reduced to the same
type of one dimensional Morse-type problem (now with $N-2$ parameters)
through a canonical transformation containing the new total momenta
$P_1=\sum_{i=1}^{N}{p_i}$.

\subsubsect{Deformed Morse systems}

\noindent $\bullet$ The {\it standard} case. The phase space
realization
\bea
&& \J_3= 2 p_1  ,\cr
&& \J_+= \tfrac12  e^{- q_1}  ,\cr
&& \J_-= - 2 \left(\Sz {p_1}\right)^2 e^{q_1} - a\,e^{q_1} ,
\label{mps}
\eea
leads to the $sl_q(2)$ Poisson algebra (\ref{lc}). 
Therefore, the one-particle Hamiltonian
\be
 H_{z}=\tfrac18 \J_3^2 - 4\,s\,\J_+ + 4\,s\,\J_+^2,
\label{msz}
\ee
does not change under deformation:
\be
H_{z}=\tfrac12 p_1^2 + s\,(e^{-2\,q_1} -2\,e^{-q_1}),
\ee
and the deformed Casimir element (\ref{le}) is just $-a/2$.

Let us now construct the associated two-body integrable deformation. By
applying the deformed coproduct onto (\ref{msz}) and with the aid of
two copies of the realization (\ref{mps}) we get
\bea
&& H_{z}^{(2)}=\tfrac12 (p_1+p_2)^2 + s\,(e^{-2\,(q_1-z\,p_2)} 
-2\,e^{-(q_1-z\,p_2)}) + \cr
&& \qquad\qquad +
s\,(e^{-2(q_2 + z p_1)} -2\,e^{-(q_2 + z p_1)})  
+ 2\,s\,e^{-\{(q_1-z\,p_2) + (q_2 + z p_1)\} }.
\label{morseh2z}
\eea
If we substitute again in terms of $P_1$, we see that $P_2$
does appear within the Hamiltonian, and the deformation would imply that
$Q_2$ is no longer a constant of motion (however, note the persistent
coupling of the type $q_1-z\,p_2$ and $q_2+z\,p_1$). The additional
integral of
motion for (\ref{morseh2z}) is obtained as the
phase space realization of 
\be
C_{z}^{(2)}=\co_{z}(C_z)=\left(\frac{\sinh (\tfrac{z}{2}
\co_{z}(\J_3))}{z}\right)^2 +
\co_{z}(\J_+)\,\co_{z}(\J_-).
\label{morsecza}
\ee 
Explicitly, (\ref{morsecza}) gives the following two-particle function:
\bea
&& \!\!\!\!\!\!\!\!\!\!\!\!\!\!\!\!\!\!\!\!
C_{z}^{(2)}=\left( \frac{\sinh (z(p_1+p_2))}{z}
\right)^2 -
\left ( \left(\frac{\sinh z\,p_2}{z}\right)^2 + \frac{a_2}{2} \right) 
\left (e^{-2 z p_1} +  e^{-(q_1-q_2) - z\, (p_1 -p_2) } \right)\cr
&& - \left ( \left(\frac{\sinh z\,p_1}{z}\right)^2 + \frac{a_1}{2} 
\right) 
\left (e^{2 z p_2} +  e^{(q_1-q_2) - z\, (p_1 -p_2) } \right).
\label{morsecz1}
\eea
Note that the limit $z\to 0$ of (\ref{morsecz1}) leads to
(\ref{morsessd}).

\noindent $\bullet$ The {\it non standard} case. The deformed phase
space realization corresponding to the $-a/2$ value of the deformed
Casimir is now
\bea
&& \J_3= 2 \frac{\sinh(\h\,e^{-q_1}/2)}{(\h\,e^{-q_1}/2)} p_1  ,\cr
&& \J_+= \tfrac12  e^{- q_1}  ,\cr
&& \J_-= - 2 e^{q_1} \frac{\sinh(\h\,e^{-q_1}/2)}{(\h\,e^{-q_1}/2)}
\,p_1^2
- \h\,\frac{a_1}{\sinh(\h\,e^{-q_1}/2)}.
\label{mpns}
\eea
Therefore, the one-particle Hamiltonian
\be
{\cal H}=\tfrac18 \J_3^2 - 4\,s\,\J_+ + 4\,s\,\J_+^2,
\label{hmns}
\ee
leads to:
\be
H_{\h}^{(2)}=\tfrac12
\left(\frac{\sinh(\h\,e^{-q_1}/2)}{(\h\,e^{-q_1}/2)}\right)^2 p_1^2 +
s\,(e^{-2\,q_1} -2\,e^{-q_1}).
\ee

The two-particle Hamiltonian is obtained from the deformed coproduct
(\ref{lr}) of  (\ref{hmns}):
\bea
&& H_{\h}^{(2)}=\tfrac14  
\left(e^{-\tfrac{\h}{2}  e^{- q_1}} 2
\frac{\sinh(\h\,e^{-q_2}/2)}{(\h\,e^{-q_2}/2)} \,p_2 +
2 \frac{\sinh(\h\,e^{-q_1}/2)}{(\h\,e^{-q_1}/2)} e^{\tfrac{\h}{2} e^{-
q_2}}\,p_1 
\right)^2
+ \cr
&& \qquad\qquad +
s\,(e^{-2\,q_1} -2\,e^{-q_1}) + 
s\,(e^{-2\,q_2} -2\,e^{-q_2}) + 2\,s\,e^{-(\,q_1+q_2)}.
\label{morseh2zns}
\eea
Again, the role of the $p_1+p_2$ dynamical variable is no longer
relevant under deformation. The two-particle Casimir would be obtained
as the phase space realization of
\be
\co_{\h}(C_\h)=\tfrac 14 (\co_{\h}(\J_3))^2 +
\frac{\sinh_{\h}(\h\,\co_{\h}(\J_+))}{\h}\,\co_{\h}(\J_-)
\label{morsecz}
\ee 
in terms of two copies of the non-standard deformation (\ref{mpns}).

\subsect{The potential $x^2 + 1/x^2$}

The following realization of $sl(2)$
\bea
&& J_3= p_1\,q_1  ,\cr
&& J_+= \tfrac12  p_1^2  + \frac{c_1}{q_1^2},\cr
&& J_-= -\tfrac 12 q_1^2 ,
\label{cent}  
\eea
underlies the $sl(2)$ dynamical symmetry of the above potential, since
by defining
\be
 H_c=J_+ - \omega^2 \,J_-
\label{hc}
\ee
we obtain that
\be
H=\tfrac12 (p_1^2 + \omega^2\,q_1^2) + \frac{c_1}{q_1^2}.
\label{centclas}
\ee
Note that the Casimir function is related to the centrifugal potential:
\be
C_c=\tfrac14 J_3^2 + J_+\,J_-=-c_1/2.
\ee

A two-particle Hamiltonian with coalgebra symmetry can be immediately
derived by computing the coproduct of (\ref{hc}). Since this dynamical
Hamiltonian is linear in the generators, we have that
\be
H^{(2)}=\tfrac12 (p_1^2 + p_2^2) + \tfrac12\,\omega^2 (q_1^2 + q_2^2) 
+ \frac{c_1}{q_1^2}+ \frac{c_2}{q_2^2},
\label{frt}
\ee
in case we assume both phase space representations not to be the same
($c_1\neq c_2$). The Casimir function for this well-known Hamiltonian
is again the coproduct of $C_c$ in the actual representation, which
reads
\be
\co(C_c)=C^{(2)}(q_1,q_2,p_1,p_2)=
- \tfrac14 (p_1 q_2 - p_2 q_1)^2- \frac{(q_1^2 +  q_2^2)}{2}
\left(\frac{c_1}{q_1^2}+ \frac{c_2}{q_2^2}  \right).
\label{cascen}
\ee

\subsubsect{Deformed $x^2 + 1/x^2$ systems}

The very same deformation machinery can be now used provided suitable
deformed realizations generalizing (\ref{cent}) are found. Hereafter it
will be useful to consider the function
\be
\Iz{x}{t}:=\frac{\sinh(t\,x/2)}{t\,x}.
\label{sz}
\ee
Note that $\lim_{t\to 0}{\Iz{x}{t}}=1/2$.

\noindent $\bullet$ The {\it standard} case. 
The deformed realization is
\bea
&& \J_3= p_1\,q_1  ,\cr
&& \J_+= \Iz{p_1\,q_1}{z} \,p_1^2+
\frac{1}{2 \,\Iz{p_1\,q_1}{z}}\,\frac{c_1}{q_1^2},\cr
&& \J_-=
-\Iz{p_1\,q_1}{z} \,q_1^2   . \label{pcents}
\eea
The one-particle Hamiltonian derived from 
\be
 H_{z}=\J_+ - \omega^2 \,\J_-
\label{hcent}
\ee
 in this realization becomes:
\be
H_{z}=\Iz{p_1\,q_1}{z}
\left(p_1^2 +
\omega^2
q_1^2\right) +
\frac{1}{2\,\Iz{p_1\,q_1}{z}}\,\frac{c_1}{q_1^2}.
\ee
The deformed Casimir function is again
\be
C_z=\left(\Sz{\J_3/2}\right)^2 + \J_+\,\J_-=-c_1/2.
\ee

When the deformed coproduct of (\ref{hcent}) is realized in terms of
two phase space realizations of the type (\ref{pcents}) leads to the
following integrable two-particle Hamiltonian:
\bea
&&\back\back H_{z}^{(2)}=\left( \Iz{p_1\,q_1}{z}
\left(p_1^2 +
\omega^2
q_1^2\right) +
\frac{1}{2\,\Iz{p_1\,q_1}{z}}\,\frac{c_1}{q_1^2}\right)
\,e^{z\,p_2 q_2/2} +
\cr && \left( \Iz{p_2\,q_2}{z}
\left(p_2^2 +\omega^2 q_2^2\right) +
\frac{1}{2\,\Iz{p_2\,q_2}{z}}\,\frac{c_2}{q_2^2}\right)
\,e^{-z\,p_1 q_1/2} \cr
&&\back\back =H^{(2)}+ z\left\{ 
p_2\,q_2\left( \tfrac12 p_1^2 + \tfrac12\,\omega^2
q_1^2 + \frac{c_1}{q_1^2} \right) -
p_1\,q_1\left( \tfrac12 p_2^2 + \tfrac12\,\omega^2
q_2^2 + \frac{c_2}{q_2^2} \right) \right\} +
o[z^2].
\label{centhz2s}
\eea
Note that this power series expansion shows how the undeformed
one-particle Hamiltonians arises in the first order of the
perturbation. As usual, the constant of the motion wil be given by
(\ref{morsecza}) where we should use two copies of the proper
realization (\ref{pcents}). As a result, we obtain
\bea
&& \!\!\!\!\!\!\!\!\!\!
C_{z}^{(2)}=\left( \frac{\sinh (z(p_1+p_2)/2)}{z}
\right)^2 \cr
&& \!\!\!\!\!\! -
\left \{  e^{- z\,p_1\,q_1 } \left( \Iz{p_2\,q_2}{z}^2\,p_2^2\,q_2^2 +
 \frac{c_2}{2} \right)
+ e^{ z\,p_2\,q_2 } \left( \Iz{p_1\,q_1}{z}^2\,p_1^2\,q_1^2 +
 \frac{c_1}{2}
\right)
\right\}\label{segcascom}\\
&& \!\!\!\!\!\! - \Iz{p_1\,q_1}{z}\,\Iz{p_2\,q_2}{z} \left\{ 
(\,p_1^2\,q_2^2 + \,p_2^2\,q_1^2) + \frac{1}{2}\left( 
\frac{c_1}{\Iz{p_1\,q_1}{z}^2}\,\frac{q_2^2}{q_1^2}+
\frac{c_2}{\Iz{p_2\,q_2}{z}^2}\,\frac{q_1^2}{q_2^2}
\right)
\right\}
\nonumber
\eea
A straightforward computation shows that the limit $z\to 0$ of this
integral gives the undeformed one (\ref{cascen}).

\noindent $\bullet$ The {\it non-standard} case. The deformed
realization for the non-standard $sl_\h (2)$ Poisson algebra with
$\J_-$ as primitive generator (\ref{lsm}) is:
\bea
&& \J_3= 2\,\Iz{q_1^2}{\h}\,q_1\,p_1  ,\cr
&& \J_+= \Iz{q_1^2}{\h}\, p_1^2 + 
\frac{c_1}{2\,\Iz{q_1^2}{\h}\,q_1^2},\cr
&& \J_-= - \tfrac12 \,q_1^2   .
\label{cenrns}
\eea
If we consider again the dynamical one-particle Hamiltonian as 
\be
H_{\h}=\J_+ - \omega^2 \,\J_-
\label{centdh}
\ee
we obtain the following deformation of (\ref{centclas}):
\be
H_{\h}=p_1^2\,\Iz{q_1^2}{\h} + \tfrac12
\omega^2\,q_1^2+ 
\frac{c_1}{2\,\Iz{q_1^2}{\h}\,q_1^2}.
\ee
Note that, as a particular feature of this case, if we consider the
dynamical Hamiltonian
$\tilde{
H_{\h}}=\left({\sinh{\h\,\J_-}}/{\,\J_-}\right)^{-1}\, H_{\h}$,
we obtain a ``natural" Hamiltonian (with the usual kinetic term $p^2$)
in which the deformation implies that the motion is not expected to be
always periodic (in fact, as long as $\h$ is increasing, non-periodic
motions become more relevant). Under such deformed realization
(\ref{cenrns}) the Casimir (\ref{lu}) turns into
$-c/2$. 

The corresponding two-body system is, as usual, provided by the
non-standard coproduct of the dynamical Hamiltonian (\ref{centdh}),
that reads
\bea
&&\back\back\back\back H_\h^{(2)}=\left( \Iz{q_1^2}{\h}
p_1^2 + 
\frac{c_1}{2\,\Iz{q_1^2}{\h}\,q_1^2}\right)\,e^{-\h\,q_2^2/2} +\cr
&&+\left(\Iz{q_2^2}{\h} p_2^2 
 + \frac{c_2}{2\,\Iz{q_2^2}{\h}\,q_2^2}\right)\,e^{\h\,q_1^2/2} +
\tfrac12
\omega^2 \,(q_1^2+q_2^2) \cr
&&\back\back =H^{(2)} + \frac{\h}{2}
\left\{q_1^2\left( \tfrac12  p_2^2  + \frac{c_2}{q_2^2} \right) -
q_2^2\left( \tfrac12  p_1^2  + \frac{c_1}{q_1^2} \right)
\right\} + o[\h^2].
\label{cent2ns}
\eea
Note that this power series expansion shows the presence of ``crossed"
oscillator and centrifugal terms coming from the deformation. Finally,
the associated constant of the motion can be deduced from the coproduct
of the non-standard Casimir (\ref{lux}) under two realizations of the
type (\ref{cenrns}).

\subsubsect{Coalgebra symmetry of the $N=2$ Calogero system}

Let us now consider the $N=2$ Calogero Hamiltonian \cite{Cal74}
\be
H_{\alpha,\beta}^{(2)}(Q,P)=\tfrac12 (P_1^2 + P_2^2) + 
\Omega^2 (Q_1^2 + Q_2^2) 
+ \frac{\alpha}{(Q_1-Q_2)^2}+ \frac{\beta}{(Q_1+Q_2)^2}.
\label{calham}
\ee
It turns out that the canonical transformation 
\be
Q_1:=(q_1 + q_2)/2,\qquad 
Q_2:=(q_2 - q_1)/2,\qquad 
P_1:=p_1 + p_2,\qquad 
P_2:=p_2 - p_1.
\label{cancal}
\ee
leads to the identification
\be
H_{\alpha,\beta}^{(2)}(q,p)=2\,H^{(2)}(q,p),
\label{clcent}
\ee
where
$H^{(2)}(q,p)$ is given by (\ref{frt}) with $\Omega^2=\omega^2/2$,
$\alpha=2\,c_1$ and $\beta=2\,c_2$.

Therefore, the Calogero
hamiltonian (\ref{calham}) does have
$sl(2)$ coalgebra symmetry, being canonically equivalent to the
(non-deformed) coproduct  ${\cal H}:=2\,(J_+ - \omega^2 \,J_-)$ and
provided two appropriate (and, in general, different) phase space
realizations of
$sl(2)$ are considered. Moreover, an integral of the motion for
(\ref{calham}) is immediately deduced from the coalgebra symmetry,
since it will be given by the (canonically transformed) coproduct of
the $sl(2)$ Casimir given by (\ref{cascen}), which takes the following
form in terms of the Calogero variables:
\be
C_{\alpha,\beta}^{(2)}(Q,P)=
- \tfrac14 (P_1 Q_2 - P_2 Q_1)^2- {(Q_1^2 +  Q_2^2)}
\left(\frac{\alpha/2}{(Q_1-Q_2)^2}+ \frac{\beta/2}{(Q_1+Q_2)^2}\right).
\label{cascencal}
\ee

As a further consequence, the coalgebra symmetry provides a systematic
procedure to get integrable deformations of the $N=2$ Calogero system.
In particular, a factor 2 times the deformed Hamiltonians
(\ref{centhz2s}) and (\ref{cent2ns}) will give rise, (respectively, and
by using the inverse of the canonical transformation (\ref{cancal})) to
the standard and non-standard deformations of
$H_{\alpha,\beta}^{(2)}(Q,P)$. Once again, corresponding
deformed integrals would be given by the inverse canonical
transformation of the coproduct of the deformed Casimirs in the
original $(q,p)$ variables.

\sect{Two-particle realizations and $2\,N$ dimensional systems}

So far we have considered one-particle phase space realizations of
coalgebras, but this is not the most general possibility. For instance,
the classical analogue of the so called Jordan-Schwinger (JS)
realization of $sl(2)$ would be 
\bea
&& J_3=a_1^+\,a_1 - b_1^+\,b_1 ={\cal N}_1^a - {\cal N}_1^b ,\cr
&& J_+= b_1\,a_1^+,\label{jsc}\\
&& J_-= a_1\,b_1^+   ,
\nonumber
\eea
where $\pois{a_1}{a_1^+}=\pois{b_1}{b_1^+}=1$. This realization can be
used to construct integrable systems by using the coalgebra approach,
but now each $sl(2)$ copy will have two degrees of freedom and we would
obtain a $2\,N$ dimensional system, whose complete integrability will
be linked to the existence of $2\,N$ quantities in involution. 

The previous formalism provides us with $N$ of them (the
$m$-th coproducts of the Casimir and the $N$-th coproduct of
the dynamical Hamiltonian). However $N$ more integrals are also
available if we take into account that, under (\ref{jsc}), the Casimir
of $sl(2)$ is no longer a numerical constant, but a two-particle
function. Namely,
\be
C_i=\tfrac14 (J_3^i)^2 + J_+^i\,J_-^i= \tfrac14 
(a_i^+\,a_i + b_i^+\,b_i)^2
=\tfrac14 ({\cal N}_i^a + {\cal N}_i^b)^2.
\label{casJS}
\ee
And we have $N$ of this quantities (in terms of classical number
operators) that, by construction, will commute with the
$N$ integrals coming from the coproduct. For instance, from
(\ref{gmg}) we find that the JS classical Gaudin magnet is 
\bea
&&\back\back\back\back H_{JS}^{(N)}=\sum_{i=1}^{N}{C_i} +
\sum_{i<j}^{N}{(\tfrac12\,J_3^i\, J_3^j +J_-^i\, J_+^j +
J_+^i\, J_-^j)}\cr
&&\back\back\back\back =\tfrac14 \sum_{i=1}^{N} {({\cal N}_i^a +
 {\cal N}_i^b)^2}
+ \tfrac12 \sum_{i<j}^{N} {({\cal N}_i^a -
 {\cal N}_i^b)({\cal N}_j^a -
{\cal N}_j^b)    }
+ \sum_{i<j}^{N} {(a_i^+\,b_j^+\,a_j\,b_i +
  b_i^+\,a_j^+\,b_j\,a_i )}.
\label{hk}
\eea
Therefore, this hamiltonian is completely integrable, since it is in
involution with both the $C_i$ functions (\ref{casJS}) and the ``lower
dimensional" Hamiltonians $H_{JS}^{(m)}$. Finally, we remark that the
last term in (\ref{hk}) is just the classical counterpart of a
longe-range interacting system with a four-wave interaction
Hamiltonian  mixing each pair of sites (see \cite{Jurco} for related
quantum optical integrable systems).

\subsect{Jordan-Schwinger map and coalgebra structure}

At this point, it becomes clear that a suitable deformation of the
classical JS map (\ref{jsc}) would give rise to a new class of $2\,N$
dimensional
$sl_q(2)$ invariant Hamiltonians. The answer to this question can be
directly related to the study of the reducibility properties of the
classical JS map, and it will lead us to a new interesting type of
coalgebra structures.

A straightforward computation shows that the $\JJ_i$ functions
\bea
&& \JJ_3= \Gco(J_3)=1 \otimes J_3 + J_3\otimes 1,\cr
&& \JJ_+= \Gco(J_+)=1 \otimes J_+ - J_+\otimes 1,\label{copg}\\
&& \JJ_-= \Gco(J_-)=1 \otimes J_- - J_-\otimes 1   .
\nonumber
\eea
close a $sl(2)$ algebra. We could now consider a pair of ``harmonic
oscillator" realizations --given by (\ref{cent}) with $c=0$-- and
obtain a realization of (\ref{copg}) on the two-particle phase space in
the form
\bea
&& \JJ_3= p_1\,q_1 + p_2\,q_2   ,\cr
&& \JJ_+= \tfrac12 ( p_1^2 - p_2^2) ,\cr
&& \JJ_-= -\tfrac 12 ( q_1^2 - q_2^2) .
\label{hosc}  
\eea
Now, the following canonical transformation can be defined
\bea
&& p_1= b_1 + \tfrac12 a_1^+,\qquad q_1= a_1 - \tfrac12 b_1^+,\cr
&& p_2= b_1 - \tfrac12 a_1^+,\qquad q_2= -a_1 - \tfrac12 b_1^+.
\label{cntr}
\eea
By substituting (\ref{cntr}) onto (\ref{hosc}) we recover (\ref{jsc}).
Therefore, the JS realization is canonically equivalent to a ``fermionic
coproduct" (\ref{copg}) of two irreps of $sl(2)$. Note that the map
$\Gco$ defined by (\ref{copg}) is not coassociative since
\be
(id\otimes\Gco)\circ\co \neq(\Gco\otimes id)\circ\Gco .  
\label{co}
\ee

The coalgebra structure allows us to perform the same trick in the
deformed case, where we can obtain the deformed realization by setting 
\bea
&& \Gco_z(\J_3) =1 \otimes \J_3 + \J_3\otimes 1,\cr  
&&  \Gco_z(\J_+) =e^{-\tfrac{z}{2}\J_3} \otimes \J_+ - \J_+\otimes
e^{\tfrac{z}{2}\J_3}; \label{lbb} \\
&&  \Gco_z(\J_-) =e^{-\tfrac{z}{2}\J_3} \otimes \J_- - \J_-\otimes
e^{\tfrac{z}{2}\J_3};\nonumber
\eea
These expressions are compatible with $sl_q(2)$ brackets (\ref{lc}) and
define a non-cocommutative and non-coassociative homomorphism. From
them, the $sl_q(2)$ generators can be expressed in terms of two phase
space realizations (\ref{pcents}) with $c=0$:
\bea
&& \tilde{f}_3=\Gco_z(\J_3) =p_1\,q_1 + p_2\,q_2 ,\cr  
&& \tilde{f}_+=\Gco_z(\J_+) =e^{-\tfrac{z}{2}p_1\,q_1 }\,
\Iz{q_2\,p_2}{z} \,p_2^2 -
\Iz{q_1\,p_1}{z}\,
\,p_1^2\, e^{\tfrac{z}{2}p_2\,q_2 };
\label{lbt} \\ 
&& \tilde{f}_-= \Gco_z(\J_-) =-e^{-\tfrac{z}{2}p_1\,q_1 } \,
\Iz{q_2\,p_2}{z} \,q_2^2
+\Iz{q_1\,p_1}{z}
\,q_1^2 \, e^{\tfrac{z}{2}p_2\,q_2 };\nonumber
\eea
Now, if we apply the canonical transformation (\ref{cntr}) onto the
functions
$\tilde{f}_i$, we shall obtain the appropriate standard deformation of
the JS map (\ref{jsc}), that can be recovered in the limit $z\to 0$.

Note that the $q$-deformation of the JS map has been treated in the
previous literature \cite{MMVZ} by making use of a deforming functional
approach
\cite{CZ,BN}. However, the discovery of its ``internal" coalgebra
structure makes it possible to give an answer in terms of the $su_q(2)$
properties. It is also worth to stress that the non-standard
deformation seems not to be compatible with such ``fermionic"
comultiplication.  

Finally, we mention that the construction of a $2\,N$
dimensional standard deformation of the Gaudin-JS system (\ref{hk}) is
straightforward by considering
$N$ copies of the deformed JS map (\ref{lbt}) and by representing
through them the
$N$-th deformed coproduct of the standard $sl_q(2)$ Casimir (\ref{le}).
The integrals of motion will be given by the $M$-th coproducts
$(M=2,\dots,N-1)$ of such deformed Casimir and by the $N$ different
functions $C_i^z$ defined by the expressions of the
$sl_q(2)$ Casimir on each copy of the JS realization. 

\bigskip
\bigskip

\noindent
{\Large{{\bf Acknowledgments}}}

\bigskip

\noindent A.B. has been partially supported by DGICYT (Project 
PB94--1115) from the Ministerio de Educaci\'on y Ciencia de Espa\~na
and by Junta de Castilla y Le\'on (Projects CO1/396 and CO2/297) and
acknowledges the hospitality during his stays in Rome.

%\footnotesize

\end{document}